\documentclass[
superscriptaddress,
nofootinbib,
nobibnotes,
notitlepage,
amsmath,amssymb,
aps,
twocolumn,
prl,
]{revtex4-1}

\usepackage{graphicx}
\usepackage[colorlinks]{hyperref}
\usepackage{bm}
\usepackage{color}
\usepackage{xspace}
\usepackage{makecell}
\usepackage{multirow}
\usepackage{dcolumn}
\newcolumntype{d}{D{.}{.}{-1}}
\usepackage{booktabs}
\usepackage{threeparttable}

\def\beq{\begin{eqnarray}}
\def\eeq{\end{eqnarray}}

\usepackage[normalem]{ulem}
\usepackage{physics}
\usepackage{ifthen}
\usepackage{aas_macros}

\newcommand{\av}[1]{\left\langle#1\right\rangle}

\newcommand{\Mpch}{h^{-1}\mathrm{Mpc}}

\begin{document}

\title{Detection of the CMB lensing -- galaxy bispectrum}
%\shorttitle{$\av{\delta_g \delta_g \kappa}$ detection}
%\shortauthors{Farren, Sherwin, Bolliet, Namikawa, Ferraro, Krolewski}

%\author{\textit{et al.}}
%\collaboration{MUSO Collaboration}%\noaffiliation

\author{Gerrit~S.~Farren}\email{gsf29@cam.ac.uk}
\author{Blake~D.~Sherwin}
\author{Boris~Bolliet}
\affiliation{DAMTP, Centre for Mathematical Sciences, University of Cambridge, Wilberforce Road, Cambridge CB3 OWA, UK}
\affiliation{Kavli Institute for Cosmology Cambridge, Madingley Road, Cambridge CB3 0HA, UK}

\author{Toshiya~Namikawa}
\affiliation{Center for Data-Driven Discovery, Kavli IPMU (WPI), UTIAS, The University of Tokyo, Kashiwa, 277-8583, Japan} 
\affiliation{DAMTP, Centre for Mathematical Sciences, University of Cambridge, Wilberforce Road, Cambridge CB3 OWA, UK}

\author{Simone~Ferraro}
\affiliation{Physics Division, Lawrence Berkeley National Laboratory, Berkeley, CA 94720, USA}
\affiliation{Berkeley Center for Cosmological Physics, University of California, Berkeley, CA 94720, USA}

\author{Alex~Krolewski}
\affiliation{Perimeter Institute for Theoretical Physics, 31 Caroline St. North, Waterloo, ON NL2 2Y5, Canada}
\affiliation{Waterloo Centre for Astrophysics, University of Waterloo, Waterloo, ON N2L 3G1, Canada}

%\correspondingauthor{Gerrit~S.~Farren}

\date{\today}

\begin{abstract} 
We present a first measurement of the galaxy-galaxy-CMB lensing bispectrum. The signal is detected at $26\sigma$ and $22\sigma$ significance using two samples from the unWISE galaxy catalog at mean redshifts $\bar{z}=0.6$ and $1.1$ and lensing reconstructions from \textit{Planck} PR4. We employ a compressed bispectrum  estimator based on the cross-correlation between the square of the galaxy overdensity field and CMB lensing reconstructions. We present a series of consistency tests to ensure the cosmological origin of our signal and rule out potential foreground contamination. We compare our results to model predictions from a halo model previously fit to only two-point spectra, finding reasonable agreement when restricting our analysis to large scales. Such measurements of the CMB lensing galaxy bispectrum will have several important cosmological applications, including constraining the uncertain higher-order bias parameters that currently limit lensing cross-correlation analyses.

%\boris{could simply call this a ``projected-field power spectrum estimator", rather than ``compressed bispectrum"?} \blake{while I agree it is kinda projected fields, for me projected fields is very much associated with a particular ksz method, so might keep as is}

%\boris{We provide a theoretical model for this observable that we implement in \href{https://github.com/CLASS-SZ/class_sz}{\texttt{class\_sz}}.}
    
\end{abstract}

\maketitle

\section{Introduction and Motivation}
Over the past decades, analyses of the power spectrum in large-scale structure have provided us with increasingly powerful constraints on cosmology. However, since structure formation becomes non-linear on small scales, a wealth of additional information about both cosmology and astrophysics is encoded in small-scale non-Gaussian statistics. In recent years, cosmologists have pioneered precise measurements of such non-Gaussian statistics in both cosmic shear and galaxy clustering observables. However, non-Gaussian statistics involving CMB gravitational lensing have generally not yet been detected, with the notable exception of one $5\sigma$ measurement of the Lyman-$\alpha$ -- CMB lensing bispectrum \citep{2016PhRvD..94j3506D}. 

Such a signal is generically expected to be present even in the absence of primordial non-Gaussianity due to the non-linear, gravitational evolution of the matter density field. Different approaches of probing non-Gaussian signals in CMB lensing analyses have been studied theoretically. Several non-Guassian statistics, for example the CMB lensing probability density function, peak counts \citep{Liu2016}, and the CMB lensing bispectrum \citep{Namikawa2018} have been discussed in the context of studying the non-linear evolution of the matter field. Furthermore, the CMB lensing bispectrum was explored as a probe of modified gravity \citep{Namikawa2018}. More recent work explored the noise biases arising when the CMB lensing bispectrum is measured from reconstructed CMB lensing maps \citep{Kalaja2023}. The lensing bispectrum has also been explored in the context of galaxy weak lensing \citep[see e.g. Refs.][and references therein]{Cooray2001,Takada2004,Munshi2020}. An analysis including the weak lensing bispectrum from the Dark Energy Survey found consistent and improved parameter constraints \citep[by 15-25\% compared to a power spectrum only analysis;][]{Gatti2022}.

%\blake{We should cite some of the papers proposing various nongaussian lensing statistics, e.g. papers by jia liu on pdf, toshiya on bispectrum of phi, and references therein. generally need more references in intro...} 

In this paper, we present the first detection of the CMB lensing -- galaxy bispectrum; our measurement employs two powers of unWISE galaxies along with one \textit{Planck} PR4 CMB lensing map. The detection significance for our measurement is perhaps surprisingly large for a first determination, with our measurement reaching a signal-to-noise ratio (SNR) of $\sim$20. We validate our detection with a number of consistency and null-tests, finding no evidence for significant systematic errors. We also compare our measurement with expectations from a simple halo model calculation.

Measurements of the CMB lensing-galaxy-galaxy bispectrum can contribute to cosmology in several ways. They can, of course, be used to directly constrain cosmological parameters, with a different dependence than the power spectrum on the amplitude of structure and matter density parameters. They can also contribute significantly to modelling and characterising galaxy populations, e.g., placing independent constraints on the parameters of an HOD model. However, perhaps the most promising application of projected CMB-lensing-galaxy bispectra is constraining higher-order bias parameters. Assuming the bispectrum can be modelled with a consistent theoretical approach (e.g., Effective Field Theory (EFT) or Hybrid EFT (HEFT)), bispectrum information on parameters such as the non-linear galaxy bias, $b_2$, can in principle break degeneracies that are a crucial limiting factor in CMB lensing cross-correlation tomography analyses. Additional information on higher-order bias parameters can be expected to allow more powerful constraints from smaller-scale galaxy cross-correlations, while reducing the impact of prior volume effects. Ref.\,\cite{Chen2021} showed significant improvements for example in constraints on the neutrino mass from the inclusion of the bispectrum between CMB lensing and galaxy surveys.

\section{Data} \label{sec:data}

In this work we employ the following datasets. The CMB lensing map is obtained from the Planck PR4 lensing analysis \citep{Carron:2022eyg}. Although the standard lensing map including optimal filtering is our baseline throughout, we also use different variations of the \textit{Planck} PR4 lensing map for null-tests. The CMB lensing signal is reconstructed up to a maximum multipole of $L_{\kappa-\rm{max}} = 2048$.

The galaxy density maps used are either the Blue or Green sample from the unWISE galaxy catalogue. The unWISE galaxy catalogue is constructed from the Wide-Field Infrared Survey Explorer (WISE) survey \citep{Wright2010}, including four years of the post-hibernation NEOWISE phase \citep{Mainzer2011,Mainzer2014}. The WISE satellite mapped the entire sky at 3.4 (W1), 4.6 (W2), 12 (W3) and 22 (W4) $\mu$m, although NEOWISE only measured in bands W1 and W2 due to a lack of cryogen necessary for the longer-wavelength bands. As a result, unWISE \citep{Lang2014,Meisner2017} is constructed only from the much deeper W1 and W2 bands. Three galaxy samples are selected from unWISE using W1-W2 colour cuts, called the Blue, Green, and Red samples, at $z \sim 0.6$, 1.1, and 1.5, respectively. These samples are extensively described in Refs.\,\cite{Krolewski2019} and \cite{Schlafly2019}. Here we only use the lower redshift samples, Blue and Green, which have substantially larger number densities. We include corrections for observational effects affecting the galaxy number density in these samples which were first derived in Ref.\,\cite{KrolewskiFerraro22}.

\section{Detection of CMB lensing-galaxy-galaxy bispectrum from the squared galaxy field}\label{sec:detection}

We use the following estimator to rapidly compute a compressed CMB lensing-galaxy-galaxy bispectrum: We square the galaxy over density, $\delta_g(\bm \hat{n})$ in real space and measure the harmonic-space cross-correlation of the $\delta_g^2$ field with the lensing convergence field $\kappa$. The overdensity is given in terms of the observed galaxy number density in the direction of $\bm \hat{n}$, $n_g(\bm \hat{n})$, and the mean number density over the entire sky $\bar{n}$ as
\beq\delta_g(\bm \hat{n}) = \frac{n_g(\bm \hat{n}) - \av{n_g}}{\av{n_g}}.\eeq

This can be compared to methods that have been used to detect the projected field kinetic Sunyaev-Zel'dovich (kSZ) \citep[see e.g. Refs.][]{PhysRevLett.117.051301,Ferraro:2016ymw,Kusiak:2021hai}. These estimators are sometimes called skew spectra and have been studied (theoretically) in the literature, for example in Ref.\,\cite{Chakraborty:2022aok}, or previous work, eg. Refs.\,\cite{Schmittfull:2014tca} and \cite{MoradinezhadDizgah:2019xun}. The resulting power spectrum, $C_L^{g^2 \kappa}$, is related to the angular bispectrum, $B^{gg\kappa}_{\ell \ell' \ell''}$ (details are shown in the appendix below).

This compressed statistic can, of course, not capture all the information in the full bispectrum. However, we note that for non-linear couplings such as the $b_2$ higher order bias term as well as the growth term in the $F_2$ matter perturbation theory kernel, this estimator is expected to be close to optimal (see e.g. Ref.\,\cite{Chakraborty:2022aok} for a detailed examination of this estimator).

\begin{figure*}
    \centering
    \includegraphics[width=\linewidth, trim=0 0.1cm 0 0.5cm, clip]{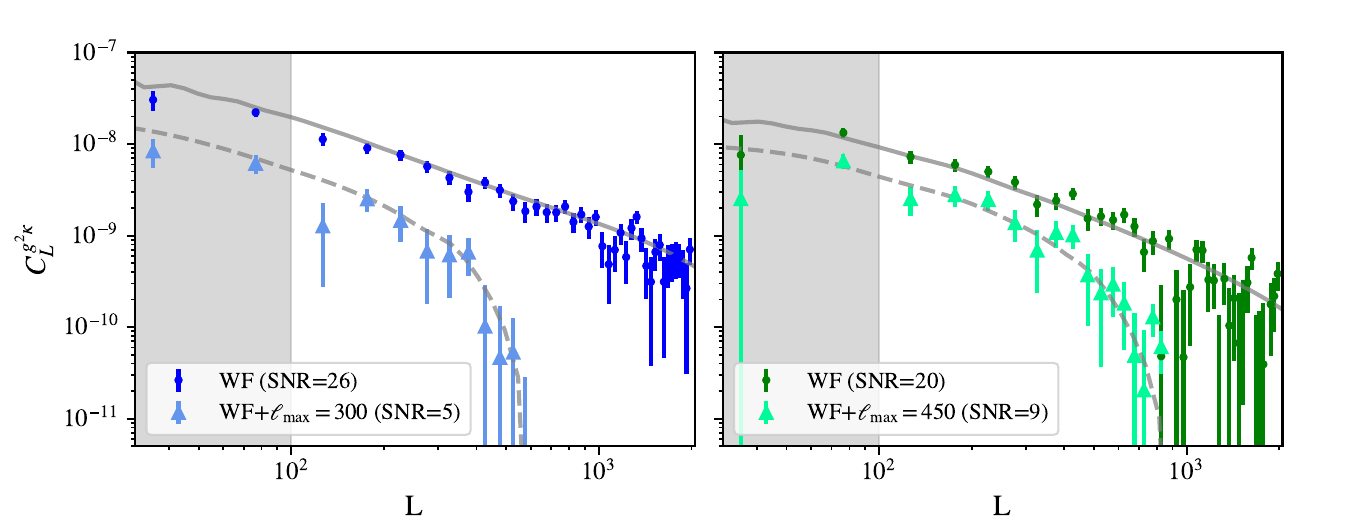}
    \caption{First detection of the galaxy-galaxy-CMB lensing bispectrum obtained via the compressed bispectrum estimator. The solid data points show a detection of $C_L^{g^2\kappa}$ at $26\sigma$ and $22\sigma$ (for $100\leq L\leq 2048$) for the Blue (\textbf{left}; $\bar{z}=0.6$) and Green (\textbf{right}; $\bar{z}=1.1$) samples of unWISE galaxies using Wiener-filtered versions of the galaxy over density maps (light, circular data points). When imposing a more restrictive cutoff on the maximum lensing multipole to conservatively guard against foreground biases, using $100\leq L \leq 1000$, we still obtain a SNR of 22 and 19 for the two samples respectively. When instead filtering the galaxy maps to exclude highly non-linear scales ($\ell<300$ and $\ell<450$ for Blue and Green respectively) we obtain $5\sigma$ and $9\sigma$ detections (on scales $100\leq L \leq 2*\ell_{\rm{max}}$; dark, triangular data points). The grey lines show an approximate, halo-model based prediction for $C_L^{g^2\kappa}$ based on the HOD fit in Ref.\,\cite{Kusiak2023} (solid for Wiener-filter and dashed for conservative scale cuts). We note that we do not perform a fit of the model parameters to the observed bispectrum and so we only compare with theory curves to check that the form of our results does not drastically differ from theoretical expectations and that a detection of the signal is unsurprising given the size of our errors. \label{fig:main_figure}}
\end{figure*}

In this work we present two different measurements of the compressed CMB lensing-galaxy-galaxy bispectrum. In order to optimise the SNR we employ a Wiener-filtered galaxy map. The Wiener-filter can be written in harmonic space as
\beq a^{g,\text{WF}}_{\ell m} = a^g_{\ell m} \frac{C_\ell^{gg, \rm{fid}}}{C_\ell^{gg,\rm{fid}}+N_\ell^{gg}}\eeq
where $N_\ell^{gg}$ is the shot noise. By default we also filter the galaxy density to remove all modes $\ell<20$ in order to suppress spurious power from large scale systematics. Furthermore, we suppress small scale fluctuations by setting all modes with $\ell\geq3000$ to zero. We do not additionally filter the lensing reconstruction in any way. With this filtering scheme the compressed bispectrum is detected at $26\sigma$ for the Blue sample and at $22\sigma$ for the Green sample. This uses all bandpowers in $C_L^{g^2 \kappa}$ for which $L\geq100$. This large scale cut is based on fluctuations observed for the lowest two bandpowers in the null-test described below\footnote{Our null-tests pass even when bandpowers with $L<100$ are included, but we consistently find lower PTEs across all tests when those large scales are included and thus conservatively exclude them.}. The maximum available value of $L$ is set by the lensing reconstruction which only includes multipoles up to $L_{\kappa-\rm{max}}=2048$. The $C_L^{g^2 \kappa}$ bandpowers are shown in Fig.\,\ref{fig:main_figure}. Alternatively, when restricting the maximum lensing multipole to $L_{\kappa-\rm{max}}=1000$ to conservatively guard against foreground systematics in the lensing reconstruction, we still obtain a SNR of 22 and 19 for the Blue and Green samples respectively. 

While the bispectrum measured from the Wiener-filtered map yields a high SNR detection of $C_L^{g^2 \kappa}$ it contains significant information from small, highly non-linear scales. Such scales are likely outside the regime of validity of the perturbative models commonly used in cosmological analysis. We thus also show results obtained using more conservative filters. We remove all modes $\ell > \ell_{g-\rm{max}} =300$ for the Blue sample and $\ell> \ell_{g-\rm{max}} =450$ for the Green sample. These scale cuts are chosen to correspond approximately to a maximum scale $k_{\rm{max}} = 0.2 \Mpch$ at the mean redshift of the two samples ($\bar{z}=0.6$ for Blue and $1.1$ for Green). For this more conservative measurement we also impose a more restrictive cut on large scales in the galaxy density, removing all modes $\ell<50$. Using these more conservative scale cuts we still obtain a detection at $5\sigma$ and $9\sigma$ for the Blue and Green samples of galaxies respectively using multipoles for which $100\leq L<2\ell_{g-\rm{max}}$ (the signal vanishes for larger multipoles, see Eq.\,\ref{eq:compressed_bispectrum}). These bandpowers are also shown in Fig.\,\ref{fig:main_figure} for comparison.

\section{Covariance} \label{sec:cov}

\begin{figure}
    \centering
    \includegraphics[width=\linewidth, trim=0 0.7cm 0cm 0cm, clip]{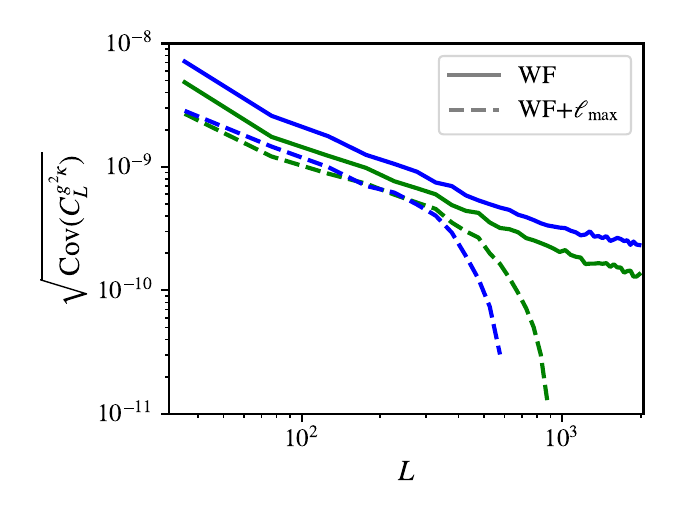}
    \caption{Diagonal elements of ${\rm Cov}(C_L^{g^2 \kappa})$. The uncertainties on the bandpowers obtained by squaring the Wiener-filtered galaxy density maps and cross-correlating it with the CMB lensing reconstruction are shown as \textbf{solid} lines. The \textbf{dashed} lines show uncertainties on our analysis with more conservative scale cuts which remove small scale information from the galaxy density maps. The blue and green curves correspond to the two unWISE samples. We note that the simulations used to obtain this estimate of the measurement uncertainty do not include the bispectrum signal and thus yield the detection covariance. Independently, the Gaussian contribution to the six-point function is expected to dominate.\label{fig:covariance}}
\end{figure}

To estimate the data covariance we rely on 480 Gaussian lensing reconstruction simulations. The generation of correlated galaxy realisations is discussed in Ref.\,\cite{farren2023atacama}. These simulations do not contain the signal we are probing in this work\footnote{We verify that we indeed obtain a signal consistent with zero when applying our pipeline to these simulations.} and thus yield the covariance under the assumption that the bispectrum vanishes as appropriate for a detection claim. The disconnected, Gaussian part of the six-point function, captured by our simulations, is expected to be the dominant contribution to the data covariance. The diagonal elements of the covariance matrix are shown in Fig.\,\ref{fig:covariance} and all off-diagonal correlations are small ($<10\%$).

\section{Data systematics test}\label{sec:sys_tests}

To establish the cosmological origin of the signal observed and to rule out potential foreground contamination we perform a series of consistency tests on the data. This includes null-tests comparing different lensing reconstruction options and different large scale masks.

Firstly, we investigate stability for different variations of the lensing reconstruction. To test for potential contamination from foregrounds such as thermal Sunyaev-Zeldovich (tSZ) clusters or the Cosmic Infrared Background (CIB) we assess the consistency of bispectrum bandpowers measured using lensing reconstructions based only on the temperature or polarisation data from \textit{Planck}. We also perform a consistency test with a lensing reconstruction performed on \textit{Planck} maps for which the tSZ contamination has been explicitly deprojected using internal linear combinations (ILCs) of different frequency maps. 

\begin{table*}[t!]
\caption{Summary of null-test we performed on the data. We provide the SNR for each of the analysis variations as well as the $\chi^2$ for the null bandpowers ($\Delta C_L^{g^2 \kappa} = C_{L, \rm variation}^{g^2 \kappa} -  C_{L, \rm baseline}^{g^2 \kappa}$) and the corresponding probability to exceed (PTE). The tests comparing different large-scale masks are not necessarily expected to pass due to fluctuations in galaxy selection properties over the survey footprint. Among null-tests expected to pass we find no failing null tests ($\rm{PTE}<0.05$). Note that the analyses using different lensing reconstructions based only on a subset of the \textit{Planck} data (TT only, EE \& EB, and TE, EE, \& EB) and those which explicitly deproject contamination from SZ clusters are not the optimally filtered \texttt{PR4} reconstructions, but instead the \texttt{PR4 2018-like} results. \label{tab:null_summary}}
\centering
\begin{ruledtabular}
\begin{tabular}{ldddddd}
     & \multicolumn3c{Blue} & \multicolumn3c{Green}\\
    & \multicolumn1c{SNR} & \multicolumn1c{$\chi^2 (\Delta C_L^{g^2\kappa})$} & \multicolumn1c{PTE} & \multicolumn1c{SNR} & \multicolumn1c{$\chi^2 (\Delta C_L^{g^2\kappa})$} & \multicolumn1c{PTE}\\
    \hline
    TT only & 25.2 & 30.6 & 0.80 & 21.3 & 31.8 & 0.62 \\
    EE \& EB & 10.5 & 45.7 & 0.18 & 10.9 & 35.0 & 0.47 \\
    TE, EE, \& EB & 13.2 & 43.4 & 0.25 & 12.0 & 36.7 & 0.39 \\
    SZ deproj. (MV) & 25.1 & 32.9 & 0.70 & 22.6 & 26.8 & 0.84 \\
    SZ deproj. (TT only) & 24.2 & 37.1 & 0.51 & 21.8 & 38.1 & 0.33 \\
    MV (\texttt{PR4 2018-like}) & 25.9 & 23.1 & 0.97 & 21.8 & 20.4 & 0.98 \\\hline
    40\% galactic mask & 21.1 & 44.5 & 0.22 & 18.9 & 47.8 & 0.07 \\
    ecliptic latitude $\beta>30^\circ$ & 19.6 & 37.5 & 0.49 & 18.3 & 55.1 & 0.02\\
    \bottomrule
\end{tabular}
\end{ruledtabular}
\end{table*}

We note, that such lensing reconstructions are not available for the optimally filtered version of the \textit{Planck} PR4 lensing reconstruction \citep{Carron:2022eyg}. Instead we use lensing reconstructions which apply the lensing pipeline from Ref.\,\cite{PlanckLensing} to the \textit{Planck} PR4 (NPIPE) CMB maps. These reconstructions are also described in Ref.\,\cite{Carron:2022eyg} and are found to be consistent with results obtained using a more optimal filtering scheme. We refer to these maps as \texttt{PR4 2018-like}. In particular we use reconstructions based on temperature data only (TT), polarisation data only (EE \& EB), the combination of polarisation data and the temperature-polarisation cross-correlation (TE, EE \& EB), and the minimum variance combination of temperature and polarisation combinations (MV). Additionally, we use temperature-only and minimum variance version of reconstructions which employ deprojection of the tSZ signal (SZ deproj. TT and SZ deproj. MV respectively). We also establish the consistency between the \texttt{PR4 2018-like} reconstruction and our baseline, optimally filtered \textit{Planck} PR4 reconstruction.

\begin{figure*}
    \centering
    \includegraphics[width=\linewidth]{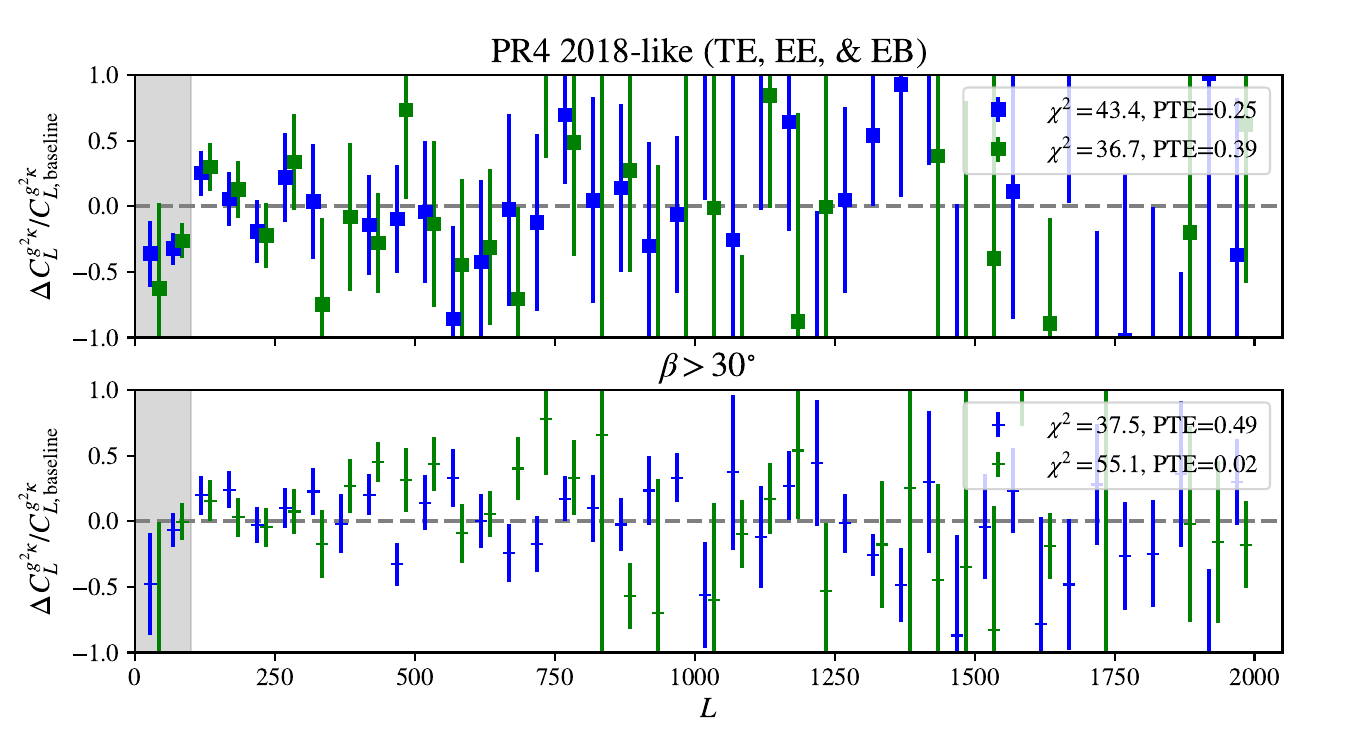}
    \caption{Null-test bandpowers for two of our tests. In both cases the bandpowers shown are $\Delta C_L^{g^2 \kappa} / C_{L, \rm baseline}^{g^2 \kappa} = C_{L, \rm variation}^{g^2 \kappa}/C_{L, \rm baseline}^{g^2 \kappa} -1$. On the \textbf{top} a null-test comparing the compressed bispectrum bandpowers obtained from our baseline analysis with an analysis using a \textit{Planck} lensing reconstruction which excludes the temperature auto-correlation when reconstructing the lensing signal. The \textbf{bottom} panel shows a null-test comparing an analysis on our baseline footprint with one restricted to ecliptic latitudes above 30$^\circ$. The latter is not necessarily expected to be consistent with zero, as variations in the WISE depth with ecliptic latitude lead to fluctuations in the galaxy selection over the survey footprint which can affect galaxy bias parameters. We do indeed find a PTE of 0.02 for the Green sample in this test. However, we detect the signal significantly in all analysis variants.\label{fig:ggk_null-tests}}
\end{figure*}

To estimate the null-test covariance matrix we obtain consistent lensing reconstruction simulations for each of the data maps used. Because the signal cancels in these null-tests the simulations are expected to yield an accurate representation of the relevant covariance. As summarised in Table~\ref{tab:null_summary}, we find the null-test bandpowers to be consistent with zero within the expected errors for all these tests (two examples are also shown in Fig.\,\ref{fig:ggk_null-tests}).

Secondly, we also perform null-tests using different galactic and ecliptic masks. As shown and discussed in Refs.\,\cite{Krolewski2019} and \cite{farren2023atacama} such tests are not necessarily expected to pass given varying selection properties for the unWISE galaxies over the survey footprint, i.e.\ the galaxy redshift distribution and bias may change in areas of deeper or shallower WISE imaging. We therefore take these test only as an approximate consistency test. We do indeed find one $\rm{PTE}<0.05$ (probability to exceed) when comparing our baseline analysis of the Green sample of galaxies and an analysis only considering ecliptic latitudes above 30$^\circ$  ($\beta>30^\circ$). The PTE for the comparison between the baseline analysis of the Green sample and measurements based on a more conservative galactic mask (40\% galactic mask) also yields a marginal PTE of 0.07. For the Blue sample of galaxies both tests are passing. Overall, while we do find statistically significant differences for some null-tests comparing the different large scale masks this is likely due to fluctuations in the galaxy selection properties. The observed bispectra are qualitatively consistent, and we emphasize that we detect the bispectrum at high significance in all analysis variants (as also shown in Table~\ref{tab:null_summary}), which strongly disfavors the possibility that our measurement comes from spurious contamination rather than a real cosmological signal.

\section{Modelling} \label{sec:modelling}

In this work we do not attempt a full cosmological analysis on the observed bispectrum. However, to demonstrate the utility of the bispectrum information we compare to theoretical predictions based on the halo model derived from the two-point function constraints on the Halo Occupation Distribution (HOD) for the unWISE samples from Ref.\,\cite{Kusiak2023}. We stress that these predictions are not a fit to any of the bispectrum measurements presented here, they are derived purely from fits to the two-point functions; they are therefore only approximate theoretical curves that are intended to allow an assessment of whether the measured spectrum's shape and amplitude agree, at some level, with expectations.

To compute the predictions we employ \texttt{class\_sz}\footnote{\href{https://github.com/CLASS-SZ/class_sz}{https://github.com/CLASS-SZ/class\_sz}} \citep{Bolliet:2023eob}. Within the halo model, the bispectrum is given as the sum of the one-, two- and three-halo contributions, $B^{gg\kappa}(k_1, k_2, k_3) = B_{1h}^{gg\kappa} + B_{2h}^{gg\kappa} + B_{3h}^{gg\kappa}$, which in turn can be computed as
\begin{widetext}
    \beq
    B_{1h}^{gg\kappa}(k_1, k_2, k_3) &=& \av{\hat{u}_{k_1}^g \hat{u}_{k_2}^g \hat{u}_{k_3}^\kappa}_n\label{eq:halo_model_terms_a}\\
    B_{2h}^{gg\kappa}(k_1, k_2, k_3) &=& \av{\hat{u}_{k_1}^g \hat{u}_{k_2}^g}_n \av{b^{(1)}\hat{u}_{k_3}^\kappa}_n P_{\rm {lin}}(k_3)+\rm{permutations}\label{eq:halo_model_terms_b}\\
    B_{3h}^{gg\kappa}(k_1, k_2, k_3) &=& 2\av{b^{(1)}\hat{u}_{k_1}^g}_n\av{b^{(1)}\hat{u}_{k_2}^g}_n\av{b^{(1)}\hat{u}_{k_3}^\kappa}_n F_2(k_1, k_2, k_3)P_{\rm{lin}}(k_1)P_{\rm{lin}}(k_2) + \rm{permutations} \label{eq:halo_model_terms_c}\\ \nonumber &&+ \av{b^{(1)}\hat{u}_{k_1}^g}_n\av{b^{(1)}\hat{u}_{k_2}^g}_n \av{b^{(2)}\hat{u}_{k_3}^\kappa}_n P_{\rm{lin}}(k_1)P_{\rm{lin}}(k_2) + \rm{permutations}.
    \eeq
\end{widetext}
Here $\hat{u}^{g}_k$ and $\hat{u}^{\kappa}_k$ are the Fourier transformed radial density profiles for galaxies and CMB lensing, $P_{\rm{lin}}(k)$ is the linear power spectrum and $F_2(k_1, k_2, k_3)$ is the standard perturbation theory tree-level bispectrum kernel. The averages, $\av{\dots}_n$, denote an average over all haloes and the permutations should be understood as permuting the galaxy and lensing profiles consistently with the associated wave numbers. The first and second order halo bias parameters, $b^{(1)}$ and $b^{(2)}$ are computed in \texttt{class\_sz} using the peak background split formulas. We refer to Ref.\,\cite{Bolliet2023} for further details on the halo-model bispectrum in this context and to Ref.\,\cite{Kusiak2023} for details on the HOD prescriptions.

Given that we aim only for an approximate, first model comparison we rely on a simple implementation that employs the Limber and flat-sky approximations. The compressed projected bispectrum is given then by
\beq
    C_L^{g^2 \kappa} &=& \frac{1}{2\pi} \int_0^\pi d\theta \int \ell^2  d\ln \ell\ w(L) w(\vert L + \ell\vert ) I(L, \ell, \theta)\nonumber\\
    \\
    I(\ell, \ell', \theta) &=& \int d\chi W_g^2(\chi) W_\kappa(\chi) B^{gg\kappa}\left(\frac{\ell'}{\chi}, \frac{\vert\bm \ell + \bm \ell'\vert}{\chi}, \frac{\ell}{\chi}\right)
\eeq
where $\vert\bm \ell + \bm \ell'\vert = \ell + \ell' + 2 \ell \ell' \cos \theta$.

The model predictions are shown as grey lines in Fig.\,\ref{fig:main_figure} (solid for the Wiener-filtered maps and dashed for the more conservative scale cuts). We find that the HOD model gives good agreement with our measurements. In the case of the analysis using Wiener-filtered galaxy maps the measurement for the Blue sample is consistent with the model prediction albeit with a relatively low PTE of 0.1. The measured $C_L^{g^2 \kappa}$ for the Green sample, while not statistically consistent (${\rm PTE}=0.0002$), also qualitatively agrees with the model prediction. The agreement between the model and our measurement is further improved for our conservative scale cuts. In this case we find the $C_L^{g^2 \kappa}$ measured for the Blue and Green samples to be fully consistent with the model prediction (${\rm PTE}=0.40$ and ${\rm PTE}=0.37$ respectively)\footnote{An initial version of this work (\href{https://arxiv.org/abs/2311.04213v1}{v1}) used the HOD constraints from Ref.\,\cite{Kusiak2022} which are obtained by fitting the power spectrum of the unWISE galaxies and their cross-correlation with \textit{Planck} lensing, but only on scales $\ell\leq 1000$. The updated model from Ref.\,\cite{Kusiak2023} is fit to smaller scales up to $\ell_{\rm max}=4000$ and yields improved agreement with our measurements.}. Given that our errorbars do not include the signal's sample variance an underestimation of the uncertainties is generically expected.

In the appendix we show the model contributions from the one-, two- and three-halo terms. As expected the analysis using Wiener-filtered versions of the galaxy density map is sensitive to relatively small scales. For the Blue sample the signal is primarily dominated by the one-halo contribution on small scales ($L\gtrsim600$). On larger scales the the two- and three-halo terms are comparable and dominate the signal. The signal from the Green sample is dominated by the one-halo contribution on small scales above $L\simeq 900$, and the three-halo term on larger scales. For the case of the more conservative scale cuts the signal primarily arises from the three-halo contribution over the entire $L$-range.

%The disagreement between measurement and model prediction is unexplained at present, but could be due to inaccuracies in the model on small scales where the two-point function provides little to no constraints on the relevant HOD parameters. Given the limited constraining power of the two-point functions for the beyond-linear-order contributions which are relevant to the bispectrum some disagreement may even be expected. This indicates that HOD constraints could be significantly improved and complemented by including bispectrum information.

The agreement between our model predictions and the measured bispectrum indicates that our measurements are consistent with the power spectrum. Because the bispectrum will generally have a different parameter dependence to the power spectrum we expect bispectrum measurements to improve on constraitns derived from the power spectrum alone. We leave a detailed investigation of these halo model constraints which can be obtained when including the bispectrum measurements for future work.

We note that a bispectrum is also produced by the post-Born correction, whose contribution is comparable to some configurations of the $\kappa$ bispectrum \citep{PrattenLewis:2016:analytic,Namikawa:2018:bispectrum-sim}. For the galaxy-galaxy-$\kappa$ bispectrum, however, the contribution from the post-Born correction is expected to be subdominant compared to the non-linear bispectrum, because the lensing and galaxy projection kernels are very different. We confirmed this assumption using an approximate analytic calculation.

\section{Conclusions}

In this paper, we have presented a first measurement of the CMB lensing -- galaxy -- galaxy bispectrum. The signal is detected at $26$ and $22$$\sigma$ significance for the two samples of unWISE galaxies. Our measurement is by far the most precise measurement of a non-Gaussian statistic involving CMB lensing. A series of null and consistency tests demonstrates that this detection is not due to any systematic contamination of our measurement. We have compared our measurement to a simple theoretical prediction based on an HOD model previously fit to two-point function measurements, finding reasonable agreement for all choices of samples and filter scales. Going forward, the measurement of such a bispectrum with ongoing and upcoming CMB and LSS surveys has the potential to contribute significantly to both cosmology and astrophysics in several ways. Perhaps the most compelling application of such a bispectrum is that its inclusion in a CMB lensing cross-correlation analysis is expected, by pinning down higher-order bias parameters, to break degeneracies that limit the cross-correlations' constraining power. However, several other applications appear similarly promising, including the derivation of direct cosmological constraints from this bispectrum and the use of the bispectrum to constrain HOD parameters and galaxy astrophysics. We hope to explore such applications of the CMB lensing -- galaxy bispectrum in future work.

\section*{Acknowledgments}

We are grateful to Julien~Carron for providing us with different variations of the \textit{Planck} lensing reconstruction and corresponding Gaussian simulations. Furthermore, we are grateful to Colin Hill for discussions on the halo model for projected-field estimators, and making us aware of the updated HOD constraints from Ref.\,\cite{Kusiak2023}. The computing for this work was performed using resources provided through the STFC DiRAC Cosmos Consortium and hosted at the Cambridge Service for Data Driven Discovery (CSD3). 

GSF acknowledges support through the Isaac Newton Studentship and the Helen Stone Scholarship at the University of Cambridge. GSF, BB, and BDS acknowledge support from the European Research Council (ERC) under the European Union’s Horizon 2020 research and innovation programme (Grant agreement No. 851274). TN acknowledges support from JSPS KAKENHI (Grant No.\ JP20H05859 and No.\ JP22K03682) and World Premier International Research Center Initiative (WPI), MEXT, Japan. SF is funded by the Physics Division of Lawrence Berkeley National Laboratory and by the U.S. Department of Energy (DOE), Office of Science, under contract
DE-AC02-05CH11231.

\subsection*{Software}

Some of the results in this paper have been derived using the \texttt{healpy} \citep{Healpix2} and \texttt{HEALPix} \citep{Healpix1} packages. We also acknowledge use of the \texttt{matplotlib} \citep{Hunter:2007} package and the Python Image Library for producing plots in this paper. Furthermore, we acknowledge use of the \texttt{numpy} \citep{harris2020array} and \texttt{scipy} \citep{2020SciPy-NMeth} packages as well as \texttt{class\_sz}\footnote{\href{https://github.com/CLASS-SZ/class_sz}{https://github.com/CLASS-SZ/class\_sz}} \citep{Bolliet:2023eob}, itself based on the Boltzman code \texttt{CLASS}\footnote{\href{https://github.com/lesgourg/class_public}{https://github.com/lesgourg/class\_public}} \citep{2011arXiv1104.2932L,2011JCAP...07..034B} and \texttt{cosmopower}\footnote{\href{https://github.com/alessiospuriomancini/cosmopower}{https://github.com/alessiospuriomancini/cosmopower}} \citep{SpurioMancini2022}.

\bibliography{bibliography,bibliography_ACTxunWISE_2pt,software_bib}
\onecolumngrid
\appendix

\section{The compressed bispectrum on the curved sky}\label{app:projected_field_bispectrum}

As a simple estimate of the CMB lensing-galaxy-galaxy bispectrum we compute the cross-correlation of the square of the galaxy overdensity, $\delta_g^2$, and the CMB lensing convergence, $\kappa$. The square of the galaxy overdensity can be expressed in harmonic space as
\begin{equation}
    \delta_g^2(\bm\hat{n}) = \sum_\ell^m \sum_{\ell'}^{m'} \delta^g_{\ell m}\delta^g_{\ell' m'} Y_{\ell}^{m}(\bm{\hat{n}}) Y_{\ell'}^{m'}(\bm{\hat{n}}),
\end{equation}
where $Y_{\ell}^{m}$ shall denote a complex spherical harmonic function.
We can the write the spherical harmonic transform of $\delta_g^2(\bm\hat{n})$ as
\begin{equation}
    \begin{split}
        (\delta_g^2)_{L M} =& \sum_{\ell, m} \sum_{\ell', m'} \delta^g_{\ell m}\delta^g_{\ell' m'} \int d\Omega Y_{\ell}^{m}(\bm{\hat{n}}) Y_{\ell'}^{m'}(\bm{\hat{n}}) \bar{Y}_L^M(\bm{\hat{n}})\\
        =& (-1)^M \sqrt{\frac{2L+1}{4\pi}} \sum_{\ell, m} \sqrt{2 \ell+1} \sum_{\ell', m'} \sqrt{2\ell'+1}\delta^g_{\ell m}\delta^g_{\ell' m'} \begin{pmatrix} \ell&\ell'&L\\0&0&0\end{pmatrix}\begin{pmatrix} \ell&\ell'&L\\m&m'&-M\end{pmatrix}.
    \end{split}
\end{equation}
It follows that the power spectrum $\av{(\delta_g^2)_{L M} \kappa_{L'M'}^*}$ is given by
\begin{equation}
    \av{(\delta_g^2)_{L M} \kappa_{L'M'}^*} = (-1)^M \sqrt{\frac{2L+1}{4\pi}} \sum_{\ell, m} \sqrt{2 \ell+1} \sum_{\ell', m'} \sqrt{2\ell'+1}\av{\delta^g_{\ell m}\delta^g_{\ell' m'} \kappa^*_{L'M'}}\begin{pmatrix} \ell&\ell'&L\\0&0&0\end{pmatrix}\begin{pmatrix} \ell&\ell'&L\\m&m'&-M\end{pmatrix}.
\end{equation}

Due to isotropy the angular bispectrum $\av{\delta^g_{\ell m}\delta^g_{\ell' m'} \kappa _{L'M'}}$, appearing above, may be written as
\beq
\av{\delta^g_{\ell m}\delta^g_{\ell' m'} \kappa^*_{L'M'}} = (-1)^{M'}\av{\delta^g_{\ell m}\delta^g_{\ell' m'} \kappa_{L'-M'}} = (-1)^{M'} B_{\ell \ell' L'}^{gg\kappa}\begin{pmatrix} \ell&\ell'&L'\\m&m'&-M'\end{pmatrix},
\eeq
leading to
\begin{equation}\label{eq:compressed_bispectrum}
    \begin{split}
        \av{(\delta_g^2)_{L M} \kappa_{L'M'}^*} =& (-1)^{M+M'}\sqrt{\frac{2L+1}{4\pi}} \sum_{\ell, m} \sqrt{2 \ell+1} \sum_{\ell', m'} \sqrt{2\ell'+1}B_{\ell \ell' L'}^{gg\kappa}\begin{pmatrix} \ell&\ell'&L'\\m&m'&-M'\end{pmatrix}\begin{pmatrix} \ell&\ell'&L\\0&0&0\end{pmatrix}\begin{pmatrix} \ell&\ell'&L\\m&m'&-M\end{pmatrix}\\
        =&\delta_{LL'} \delta_{MM'} \sum_{\ell,\ell'} \sqrt{\frac{(2\ell+1)(2\ell'+1)}{4\pi(2L+1)}} B_{\ell \ell' L}^{gg\kappa}\begin{pmatrix} \ell&\ell'&L\\0&0&0\end{pmatrix}.
    \end{split}
\end{equation}

\section{Halo model contributions by one-, two-, and three-halo terms} \label{app:halo_model_contributions}
\begin{figure*}
    \centering
    \includegraphics[width=\linewidth]{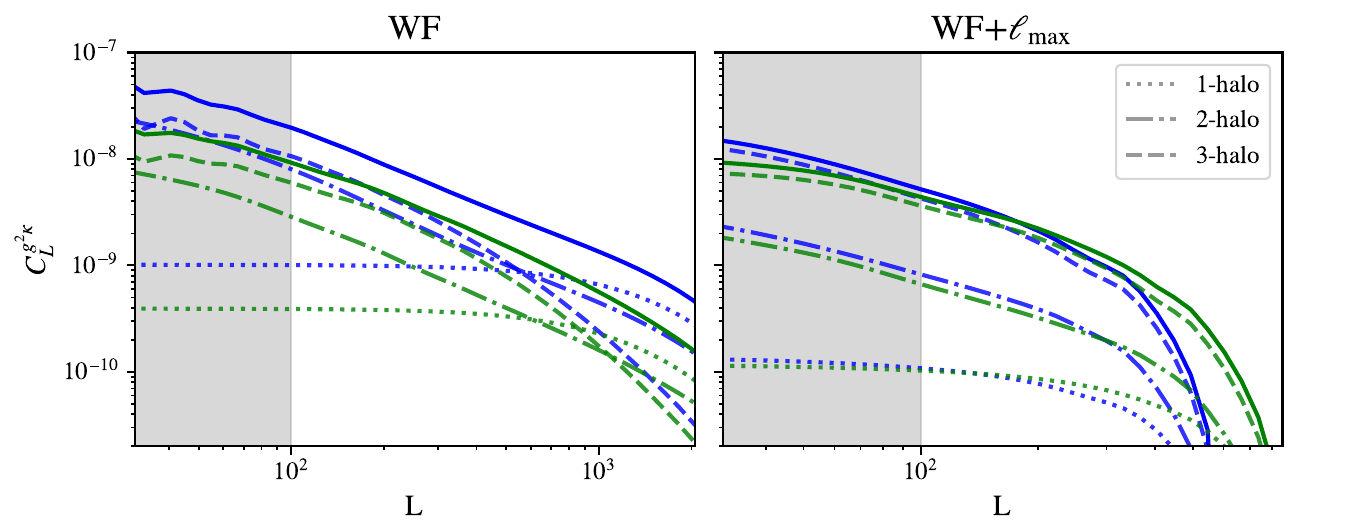}
    \caption{The figure shows the one-, two-, and three-halo contributions to the halo model prediction for the compressed galaxy-galaxy-CMB lensing bispectrum, $C_L^{g^2\kappa}$. The \textbf{left} panel shows a prediction for the case where the galaxy density maps are Wiener-filtered, while the \textbf{right} panel shows a prediction for the case in which small scales ($\ell>300$ and $\ell>450$ for the Green and Blue samples respectively) have been removed from the galaxy density maps prior to squaring them. In the former case, the information is dominated by the one halo contribution over a large range of scales, while the three-halo contribution is the dominant source of signal in the latter case. \label{fig:ggk_model_contributions}}
\end{figure*}

In Fig.\,\ref{fig:ggk_model_contributions} we show the contributions to the halo model prediction for the compressed galaxy-galaxy-CMB lensing bispectrum from the one-, two-, and three-halo terms (see Eqs.\,\ref{eq:halo_model_terms_a}-\ref{eq:halo_model_terms_c}). As above this computation assumes the two-point function based halo model constraints from Ref.\,\cite{Kusiak2023} and is not a fit to any of our bispectrum measurements. For the SNR optimised analysis (WF) the one-halo term dominants above $L\simeq600$ for the lower redshift, Blue sample and above $L\simeq900$ for the Green sample, leading to high sensitivity to the satellite distribution and the mass profile within halos. On larger scales the three-halo contribution dominates the signal from the Green sample, while two- and three-halo contributions are comparable for the Blue sample. When removing small scale information by filtering out small scales, however, the signal is dominated over the entire range of $L$s (for which the signal does not vanish) by the three-halo contribution.

\end{document}